# Time dependence of the behaviour of silicon detectors in intense radiation fields and the role of primary point defects[1]

Sorina Lazanu,[a*] Ionel Lazanu[b]

[a]*National Institute for Materials Physics, POBox MG-7, Bucharest-Magurele, Romania*

[b]*University of Bucharest, Faculty of Physics, POBox MG-11, Bucharest-Magurele, Romania*

**Abstract**

The bulk displacement damage in the detector, produces effects at the device level that limits the long time utilisation of detectors as position sensitive devices and thus the lifetime of detector systems. So, the prediction of time behaviour of detectors in hostile radiation environments represents a very useful tool. In this contribution we predict the time degradation of silicon detectors in the radiation environments expected in the LHC machine upgrade in luminosity and energy as SLHC, for detectors fabricated from silicon crystals obtained by different growth technologies, in the frame of the model developed by the authors, and which takes into account the contribution of primary defects.

## 1. Overview of the changes induced by irradiation in the operational silicon detector parameters

The studies related to irradiation effects in semiconductors, and in particular in silicon, have a long history. A vast amount of data was collected trying to understand the principal mechanisms by which the detector performances change during irradiation and deteriorate the optimised device parameters. The main operational parameters that degrade detection characteristics of the device are: the full depletion voltage $V_{dep}$, the leakage current

---

and the charge collection efficiency (CCE). These changes are due to bulk displacement damage in the lattice, which produces primary point defects, which, during annealing processes, interact between them or with impurities and produce new defects. Consequently, an increase of the leakage current in the detector is produced, a decrease of the initially satisfactory Signal/Noise ratio, and an increase of the effective carrier concentration (and thus of the depletion voltage), which ultimately increases the operational voltage of the device beyond the breakdown voltage.

In silicon detectors, the increase of the leakage current during and after irradiation is due to the generation of electron-hole pairs on the energy levels of the produced defects, an important contribution to the leakage current having those with energy levels in the vicinity of the mid-gap, and with high cross sections for carrier capture. It has been calculated in agreement with the Shockley – Read - Hall model. In the field region, the occupancy of the energy levels is determined by the balance of trapping and emission of charge carriers. The effective carrier concentration, $N_{eff}$ represents the absolute value of the difference between ionised donors and acceptors.

## 2. The model for silicon degradation

The model of kinetics of defects used in the present work has been developed previous by the authors, see for example [1] and references cited therein. It describes, at the microscopic level, the formation and evolution of defects in silicon during and after irradiation and correlates these effects with the macroscopic parameters of the device.

The incident particle interacts with the semiconductor material. The peculiarities of the interaction mechanisms are explicitly considered for each type of particle and kinetic energy. When an incident particle is slowed down in silicon, it produces different types of damage. While ionisation is the basis of particle detection and is reversible, displacement effects produce defects in the lattice – which are of interest for this analysis. The processes by which the recoil nuclei resulting from these interactions lose their energy in the semiconductor lattice are modelled. The kinetic energy dependence of the concentration of primary defects on unit particle fluence (CPD) is calculated for each incident particle and energy, and it is used in the evaluation of the generation rate of defects during irradiation in accord with Lindhard theory. In the next step is modelled the formation and time evolution of complex defects, associations of primary defects or of primary defects and impurities. The primary point defects considered here are vacancies (as "classic" vacancies and fourfold coordinated vacancy defects: $Si_{FFCD}$) and silicon self interstitials. In what regards the $Si_{FFCD}$ defect, its existence was predicted by Goedecker and co-workers [2] and its characteristics were indirectly established by Lazanu and Lazanu [3]. We established that this defect is produced simultaneously with the "classic" vacancy, with a concentration of about 10% from all vacancies per act of interaction, is stable in time, it is uniformly introduced in the bulk during irradiation and has deep energy level(s) in the gap, at least one energy level between $E_c$ – (0.46 ÷ 0.48) eV. The primary defects are introduced with a rate that is sum of contributions from thermal generation and generation by irradiation.

In the forthcoming analysis, only phosphorus, carbon and oxygen impurities are considered as pre-existent in silicon. The kinetics of defects was modelled applying the theory of diffusion-limited reactions. The complete reaction scheme for the formation and evolution of defects could be found in reference [1]. The time evolution of defect concentrations for primary and complex defects produced in silicon is the solution of the associated system of simultaneous differential equations, which do not have an analytic solution. The model is able to calculate, without free parameters, absolute values of microscopic and macroscopic quantities.

## 3. Predictions of the behaviour of detectors in radiation fields

The comparison of model calculation with experimental data at the macroscopic level has been performed. In this respect, three classes of results will be discussed here: a) related to the time evolution of macroscopic characteristics of detectors after irradiation, b) the fluence dependence of these characteristics for different technologies used, and c) predictions of detector behaviour in continuous irradiations regime as expected to SLHC for FZ and DOFZ technologies for silicon detectors.

**a)** The modifications induced by irradiation in the leakage current and effective carrier concentration of silicon detectors have been previously calculated and compared with available experimental data on detector characteristics after irradiation with electrons, protons, positive and negative pions, and neutrons (see Ref. [3]). A good agreement between model calculations and experimental data has been obtained.

**b)** The standard growth techniques of the silicon crystal utilised for detectors fabrication are Czochralski (Cz) and Float Zone (FZ). They differ mainly by the oxygen concentration and by the resistivity. Due to the believed positive role of oxygen, a technique for its incorporation into FZ silicon has been developed and DOFZ has been obtained.

Model calculations have been compared with experimental data for fluence dependence of $N_{eff}$ after pion and proton irradiation. In Figure 1, this comparison is presented for silicon irradiated with 190 MeV/c pions, where the silicon wafer was obtained using FZ, DOFZ, and Cz techniques respectively.

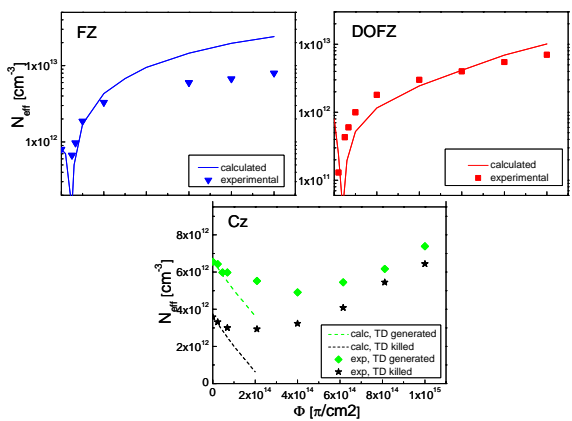

Fig 1 Fluence dependence of $N_{eff}$ for DOFZ, FZ and Cz silicon detectors irradiated with pions

It could be seen that, while for FZ and DOFZ growth methods a reasonable agreement has been obtained; in the case of Cz silicon, this agreement is restricted only to the beginning of irradiation, i.e. to low fluences. In our opinion, this fact could be attributed the presence of defects associated with oxygen in the material, and consequently supplementary processes they are taking part in during and after irradiation must be added to the reaction scheme.

For high resistivity silicon (2 kΩcm and 15 kΩcm respectively) irradiated with 24 GeV/c protons, a good agreement between model calculations and experimental data from Ref. [4] is obtained both for FZ and DOFZ material, as could be seen in Figure 2. These results show that the degradation at the highest fluences studied here is not clearly correlated with oxygen concentration in silicon or with the resistivity of the starting material. Consequently, for high fluences, it is not possible to make a choice between FZ and DOFZ materials.

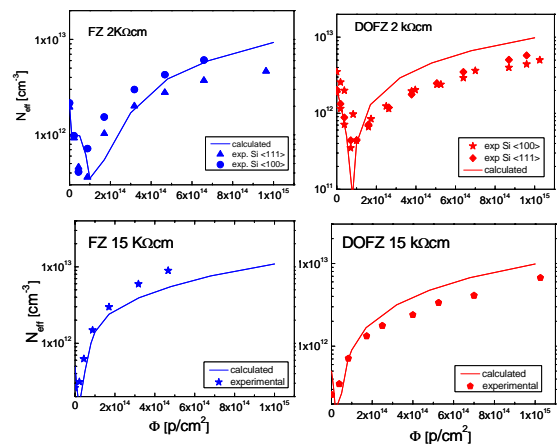

Figure 2 Fluence dependence of the effective carrier concentration of FZ and DOFZ silicon detectors irradiated with protons

**c)** In the environments considered in the present study, the detector systems will work usually between 5 to 10 years, in a continuous irradiation field. At the present time the radiation fields in approximate experimental configurations are estimated only for LHC.

For LHC upgrades, in the absence of detailed studies, only suppositions are possible. In this contribution, only the SLHC option will be discussed.

Following the idea exposed by F. Gianotti in Ref. [5], we supposed the following conditions for the collider upgrade: the luminosity is increased with an factor of ten in report with LHC, the beam energy is increased with a factor of two, and the energetic distributions of pions and protons are the same as for LHC conditions, but with the average energy of the spectra shifted to higher energy with 50 MeV. In the LHC conditions (considering for the concrete discussions the particular case of the CMS experiment), after the primary interaction, hadrons are the predominant particles in the tracker, especially low energy charged pions and protons. The maximum in the rates of primary defects generated by pions comes from the region around 200 MeV while for protons the major contribution comes from the lowest energy region. The rates of generation of defects (induced by pions and protons) are: $6.8 \times 10^8$ VI/cm$^3$/s in the LHC conditions and $7.2 \times 10^9$ VI/cm$^3$/s for SLHC hypothesis. The contributions coming from proton spectra represent about 8 % from all primary defects.

The behaviour of FZ and DOFZ silicon detectors in the fields estimated for LHC and SLHC is presented as time dependence of the density of leakage current and as effective carrier concentration, in conditions of continuous irradiation – see Figure 3. For leakage currents, the differences between DOFZ and FZ silicon are very small and could be observed especially for LHC rates in the first months of operation, where DOFZ silicon seems radiation harder.

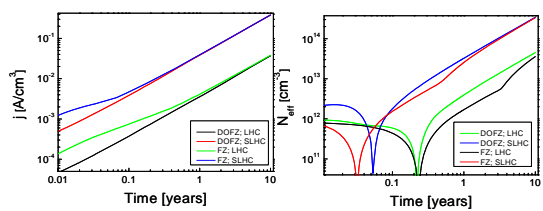

Figure 3 Time dependence of the volume density of the leakage current and effective carrier concentration for FZ and DOFZ detectors, at LHC and SLHC respectively.

As in the previous case, one could see that the benefits of oxygenation are lost, and roughly the same results are estimated for both materials, the effects being correlated with the luminosity and not with the characteristics of particle spectra.

## 4. Summary

The model is able to reproduce the main characteristics of the available experimental data on time and fluence dependence of leakage current and effective carrier concentration for FZ and DOFZ silicon. It is able to predict the behaviour of Si detectors obtained from crystals grown by these techniques in different radiation fields.

In the scenarios for radiation environments at SLHC, the produced degradation of silicon detectors scales with luminosity, and is roughly independent on particle spectra. In these conditions, there are no significant differences between FZ and DOFZ silicon. In conditions of continuous long time irradiation at high rates of generation of defects, the contribution of primary defects is important and must be considered, but these predictions need experimental confirmation. Because the processes have temperature dependence, the decrease of the temperature is strongly required, in order to diminish these macroscopic effects.

### Acknowledgments

This work has partially been supported by the Romanian Scientific National Programmes CERES and MATNANTECH under contracts C4-69/2004 and 219 (404) /2004 respectively..